\documentclass[aps,prl,reprint,twocolumn,superscriptaddress]{revtex4-2}
\usepackage[title]{appendix}
\usepackage{hyperref}
\hypersetup{colorlinks=true,linkcolor=black,citecolor=brown,filecolor=black,urlcolor=black}
\usepackage{textcomp}
\usepackage{graphicx}
\usepackage{amsmath}
\usepackage{dcolumn}
\usepackage{xcolor}

\newcolumntype{d}[1]{D{.}{.}{#1}} 

\newcommand{\V}[1]{\mathbf{#1}} 
\newcommand{\T}[1]{\texttt{#1}} 
\newcommand{\xhat}{\mbox{$\hat{\mathbf{x}}$}} 
\newcommand{\yhat}{\mbox{$\hat{\mathbf{y}}$}} 
\newcommand{\zhat}{\mbox{$\hat{\mathbf{z}}$}} 

\newcommand\Alfven{Alfv\'en }

\newcommand{\figref}[1]{Fig.~\ref{#1}}

\newcommand{\eqnref}[1]{Equation~(\ref{#1})}
\newcommand{\tabref}[1]{Table~(\ref{#1})}

\begin{document}

\title{Extended magnetic reconnection in kinetic plasma turbulence}
\author{Tak Chu Li}\email{tak.chu.li@dartmouth.edu}
\author{Yi-Hsin Liu}
\affiliation{Department of Physics and Astronomy, Dartmouth College, Hanover, NH 03755, USA}
\author{Yi Qi}
\affiliation{Laboratory for Atmospheric and Space Physics, University of Colorado Boulder, Boulder, CO 80303, USA}
\author{Muni Zhou}
\affiliation{School of Natural Sciences, Institute for Advanced Study, Princeton, NJ 08544, USA}


\begin{abstract}

Magnetic reconnection and plasma turbulence are ubiquitous processes important for laboratory, space and astrophysical plasmas. Reconnection has been suggested to play an important role in the energetics and dynamics of turbulence by observations, simulations and theory for two decades. The fundamental properties of reconnection at kinetic scales, essential to understanding the general problem of reconnection in magnetized turbulence, remain largely unknown at present. Here we present an application of the magnetic flux transport method that can accurately identify reconnection in turbulence to a three-dimensional simulation. Contrary to ideas that reconnection in turbulence would be patchy and unpredictable, highly extended reconnection X-lines, on the same order of magnitude as the system size, form at kinetic scales. Extended X-lines develop through bi-directional reconnection spreading. They satisfy critical balance characteristic of turbulence, which predicts the X-line extent at a given scale. These results present a picture of fundamentally extended reconnection in kinetic-scale turbulence.



\end{abstract}

\maketitle


Magnetic reconnection and plasma turbulence are ubiquitous in the universe. Turbulence transfers energy from large scales to small scales where the energy is dissipated. Reconnection converts magnetic energy into plasma flow and thermal energy. They are thought to be energetically and dynamically important for a range of systems, including laboratory devices, Earth's magnetosphere, the solar wind and solar corona \citep{Yamada:2010,Zimbardo:2010,Bruno:2013}, the interstellar medium and galaxy clusters \citep{zweibel09a,Goldreich:1995,Elmegreen:2004,Treumann:2015}. Reconnection has been suggested to play an important role in the energetics and dynamics of turbulence by observations, simulations and theory for decades, by dissipating turbulence energy \citep{Dmitruk:2004,Sundkvist:2007,Osman:2011,Markovskii:2011,Perri:2012a,Wan:2012,TenBarge:2013a,Zhdankin:2013,Shay:2018,Rueda:2021,Stawarz:2022,Franci:2022} and mediating the turbulent cascade \citep{Cerri:2017b,Loureiro:2017a,Boldyrev:2017,Mallet:2017a,Franci:2017,Mallet:2017b,Loureiro:2017b,Vech:2018a,Stawarz:2019,Manzini:2023}. 
The general problem of reconnection in magnetized turbulence is a field of extensive research, particularly in large-scale systems \citep{Lazarian:2020}. Here we focus on the small-scale limit of the problem, where fundamental properties of reconnection are largely unknown.

In the heliosphere, reconnection has been observed in large-scale current sheets, close to interplanetary coronal mass ejections (ICMEs) \citep{gosling05a,Phan:2009}, and reported to be extended over 10$^4$ ion gyroradii $\rho_i$ \citep{phan06a,Phan:2009,Eastwood:2021}. At kinetic scales (sub-ion scales of $k_\perp\rho_i >$ 1), recent Wind and Parker Solar Probe observations have revealed an abundance of kinetic-scale ($\simeq$ 1 $\rho_i$) current sheets near Earth and near Sun, with a scale dependence consistent with generation by a turbulent cascade \citep{Vasko:2022,Lotekar:2022}; the detection of reconnection at kinetic scales is ultimately limited by the resolution of the instruments. At electron scales, electron reconnection without coupling to ions has been recently detected by the Magnetospheric Multiscale (MMS) mission in Earth's turbulent magnetosheath \citep{Phan:18}. Three-dimensional (3D) kinetic simulations indicate patchy electron reconnection, with extents limited to $\sim$10 electron gyroradii \citep{Pyakurel:2021}. The spatial distribution of reconnection in kinetic-scale turbulence, where energy is dissipated, and the underlying physics are currently unknown. Investigating these fundamental properties of reconnection is important for understanding the general problem of reconnection in magnetized turbulence.


Identifying reconnection in turbulence is an essential step. In simulations and observations, Alfv\'enic ion or super-Alfv\'enic electron outflow jets have been used as a reconnection signature. However, outflow jets can be distorted or suppressed by turbulent flows at kinetic scales \citep{Bessho:2020,Li:2021}. In simulations, the saddle point method that defines a topological X-line has been applied, but shown to detect X-lines that are not actively reconnecting \citep{servidio09a,Servidio:2010,Wan:2014,Haggerty:2017}. Indicators based on strong currents and/or fast flows \citep{Zhdankin:2013,Rueda:2021,Sisti:2021} and the E$\times$B velocity \citep{Lapenta:2021,Pongkitiwanichakul:2021} have also been considered, but the former may not be directly related to reconnection while the E$\times$B velocity is not applicable to nonideal regions where plasma and magnetic field motions decouple.




{\it Magnetic flux transport.} 
Recently, a novel method based on magnetic flux transport (MFT), which is inherent to reconnection, has been considered in simulations and observations of plasma turbulence \citep{Li:2021,Qi:2022}. This method is based on the definition of reconnection as the transport of magnetic flux across magnetic separatrices that intersect at an X-line \citep{Vasyliunas:1975}. It measures signatures of active reconnection in the in-plane velocity of magnetic flux and its divergence, $\mathbf{U}_\psi$ and $\nabla\cdot \mathbf{U}_\psi$. Evidence for converging inward and diverging outward MFT flows at an X-line in either of the quantities provides signature of active reconnection.

$\mathbf{U}_\psi$ was derived in two dimensions (2D) using a 2D advection equation of magnetic flux \citep{YHLiu:18c,YHLiu:16}, and was later simplified and adapted for application in 3D \citep{Li:2021}, given by:
\begin{eqnarray}
 \mathbf{U}_\psi = \frac{c\, \delta E_{z}}{\delta B_p}(\zhat\times\delta\hat{b}_p),
\label{eq:Upsi2}
\end{eqnarray}
where $\delta E_z$ is the component of the fluctuating electric field parallel to the background magnetic field, and $\delta\hat{b}_p \equiv \delta \mathbf{B}_p/\delta B_p$ is the unit vector of the perpendicular or in-plane magnetic field fluctuations $\delta \mathbf{B}_p \equiv \delta B_x\xhat + \delta B_y\yhat$. 
$\mathbf{U}_\psi$ can be decomposed into in-plane electron flow and a slippage term that depends on a nonideal electric field \citep{YHLiu:18c,YHLiu:16}, discussed in \citep{Li:2021}. See also a comparison of $\mathbf{U}_\psi$ and the E$\times$B velocity in supplementary material, which includes Refs. \citep{Vasyliunas:1972,Pontin:2022}. 



The MFT method has been demonstrated to accurately identify reconnection in 2D gyrokinetic and 3D shock turbulence simulations \citep{Li:2021,Ng:2022}. Recent MMS observations have further demonstrated the accuracy of MFT statistically, having directly measured MFT signatures for active reconnection throughout Earth's magnetosphere \citep{Qi:2022}. In this Letter, we apply MFT to a 3D simulation of gyrokinetic turbulence, and present first evidence for spatially extended reconnection in kinetic-scale turbulence.


\begin{figure*}[t!]
\resizebox{6.7in}{!}{\includegraphics[scale=1.,trim=1.cm 0.1cm 0.2cm 0.3cm, clip=true]{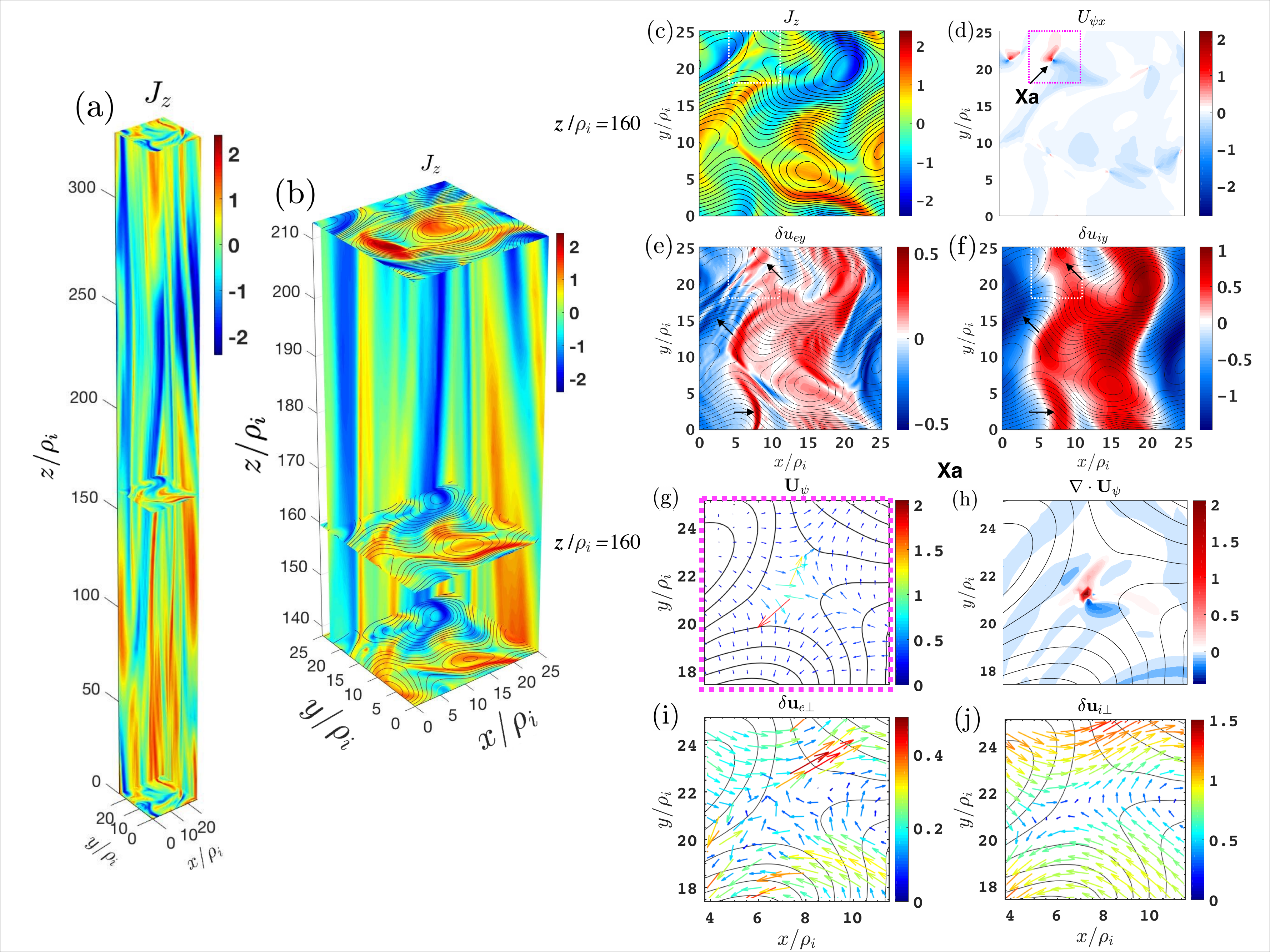}}
\caption{ \label{fig:f1}{\sffamily Reconnection identification: the parallel current density $J_z$ in (a) the 3D domain, and (b) a central region of $z/\rho_i$=140--210, overlaid with contours of the parallel vector potential $A_\parallel$. Quantities shown on the $z/\rho_i$=160 plane are (c) $J_z$, (d) the $x$-component of the MFT velocity $U_{\psi x}$, (e-f) the $y$-component of the fluctuating electron and ion bulk flow velocities, $\delta u_{ey}$ and $\delta u_{iy}$. In an enlarged region around Xa, denoted by a magenta dashed box in (d), shown are (g) vectors of $\mathbf{U}_\psi$, (h) $\nabla\cdot \mathbf{U}_\psi$, and (i-j) vectors of the fluctuating in-plane electron and ion flow velocities, $\delta \mathbf{u}_{e\perp}$ and $\delta \mathbf{u}_{i\perp}$. }
}
\end{figure*}




{\it Simulation.} The simulation was performed \citep{Li:2016} using the Astrophysical Gyrokinetics Code, \T{AstroGK} \citep{Numata:2010}. Here we specifiy a 3D generalization \citep{Li:2016,Li:2019} of the classic 2D Orszag-Tang Vortex problem \citep{Orszag:1979}. This setup consists of counterpropagating \Alfven waves along the background magnetic field $\V{B}=B_0\zhat$. More information is given in supplementary material, which includes Refs. \citep{Howes:2008a,Howes:2011a,TenBarge:2012a,Frieman:1982,Howes:2006,Abel:2008,Barnes:2009,Politano:1989,Mininni:2006,Parashar:2014b}.

To follow the turbulent cascade from the inertial range ($k_\perp\rho_i\ll 1$) to below electron scales ($k_\perp\rho_e >1$) \citep{TenBarge:2013a,TenBarge:2013b}, we specify a reduced mass ratio, $m_i/m_e=25$, which, in a simulation domain of $L_{\perp}=8\pi\rho_i$ and dimensions $(n_x,n_y,n_z)=(128,128,32)$, enables us to resolve a dynamic range of $0.25\le k_\perp\rho_i\le 10.5$, or $0.05\le k_\perp\rho_e\le 2.1$. Plasma parameters are $\beta_i=8 \pi n_i T_{0i}/B_0^2=0.01$ and $T_{0i}/T_{0e}=1$. Length, time and velocity are normalized to $\rho_i\equiv v_{ti}/\Omega_{ci}$, where $\Omega_{ci}\equiv eB_0/m_ic$, domain turnaround time $\tau_0\equiv L_\perp/\mbox{z}_0$ and electron thermal speed $v_{te}\equiv\sqrt{2T_{0e}/m_e}$. Ion velocity is instead normalized to $v_{ti}\equiv\sqrt{2T_{0i}/m_i}$. $\tau_0$ can be converted to the inverse ion gyro-frequency, a relevant time scale for reconnection, by $\tau_0$=25$\Omega_{ci}^{-1}$. The divergence of velocity is normalized to $v_{te}/\rho_e=\Omega_{ce}$.

{\it MFT application.} There are two conditions for applying MFT: (i) $k_\parallel \ll k_\perp$ and (ii) quasi-planar reconnection \citep{Li:2021}. $k_\parallel \ll k_\perp$ is consistent with anisotropic turbulence theory \citep{Cho:2004,Schekochihin:2009} and observations of solar wind and magnetosheath turbulence \citep{Alexandrova:2008b,Alexandrova:2009,Sahraoui:2013b,Chen:2017}. Quasi-planar reconnection, which is a basis for the local current sheet (LMN) coordinate widely adopted in space observations, is consistent with observations of large-scale current sheets in the solar wind (e.g. \citep{phan06a,Phan:2009}) and magnetotail turbulence (e.g. \citep{Ergun:2022}) and small-scale current sheets in the magnetosheath (e.g. \citep{Phan:18,Stawarz:2022}).


The conditions for applying MFT are well satisfied in the simulation. (i) $k_\parallel \ll k_\perp$ is observed in the system, as expected for anisotropic turbulence. (ii) The perpendicular magnetic fluctuations dominate over parallel fluctuations, {\it i.e.}, $\delta B_\parallel \ll \delta B_\perp$. Reconnection is dominated by perpendicular fluctuations, making reconnection quasi-planar. The background (guide) magnetic field also puts reconnection in the strong-guide-field limit, with a guide field $B_0 \sim$10 times the reconnection magnetic field $\delta B_\perp$.

In applying MFT, as a practical step, we add a 1\% offset to $\delta B_p$ in \eqnref{eq:Upsi2}, similar to previous work \citep{Li:2021}, such that the amplitude at the X-line (where MFT is not applicable since a source or sink term, representing flux generation or annihilation at the X-line, is not included in the advection equation) resembles those in the vicinity of the X-line. For the range of 0.01--1\% offsets, the amplitudes of $\mathbf{U}_\psi$ and $\nabla\cdot \mathbf{U}_\psi$ only vary by a factor of 2.

In identifying reconnection, MFT currently does not distinguish between ion-coupled or electron-only reconnection. Both forms of reconnection can occur in kinetic turbulence (e.g.,\citep{Franci:2022,Stawarz:2022}).


{\it Reconnection Identification.} 
We first demonstrate how MFT identifies reconnection in 3D. We show in \figref{fig:f1}(a) the parallel current density $J_z$ in the 3D domain at $t/\tau_0$ = 0.34, a time of strong reconnection activity and strong energy dissipation \citep{Li:2016}. A turbulent cascade at kinetic scales of $k_\perp\rho_i>$ 1 has also developed. Here $k_\perp$ is the perpendicular wavenumber based on the radius of flux ropes undergoing reconnection. In panel (b) $J_z$ in a central region of $z/\rho_i$=140--210 shows fine-scale structures, including small-scale current sheets, on several $xy$ planes. We first focus on the plane at $z/\rho_i$=160, and show how MFT identifies reconnection.

On the $z/\rho_i$=160 plane, shown are (c) $J_z$ and (d) $U_{\psi x}$, the $x$-component of $\mathbf{U}_\psi$. Multiple flux ropes are evident in $J_z$. $U_{\psi x}$ reveals prominent MFT flows from the two strongest X-lines. The strongest X-line, Xa, forms from flux rope merging, with the direction of inflowing flux ropes (inflow direction) primarily directed along $\xhat$. The $x$-component of $\mathbf{U}_\psi$ shows converging inflows of magnetic flux at Xa. The outflow direction is primarily directed along $\yhat$. The plasma outflow jets can be seen in the $y$-component of the fluctuating electron and ion bulk flow velocities, shown in (e) $\delta u_{ey}$ and (f) $\delta u_{iy}$. In (e), $\delta u_{ey}$ shows bi-directional electron outflow jets from Xa (arrowed), including an upward jet through the periodic boundary at $y/\rho_i$=25 appearing at the bottom left. In (f) $\delta u_{iy}$, broad ion outflow jets form. The plasma outflow jets are more broadly distributed from the X-line than the localized MFT flows.

Quantities in the enlarged region around Xa, denoted by a magenta box in (d), are shown in \figref{fig:f1}, panels (g)--(j). The vectors of $\mathbf{U}_\psi$ (shown in panel (g)) reveal clear inflows and outflows of MFT as a signature of active reconnection \citep{Li:2021}. The divergence of MFT, $\nabla\cdot \mathbf{U}_\psi$ (shown in panel (h)), shows strong localized positive and negative peaks at Xa, representing diverging outflows and converging inflows of MFT, similarly signifying active reconnection. It also has a quadrupolar structure observed in 2D \citep{Li:2021}. The MFT signatures in this 3D simulation are similar to the 2D case, although more irregular, as would be expected in 3D turbulence. 
While $\mathbf{U}_\psi$ is normalized to the electron thermal speed, when
renormalizing to the upstream electron Alfv\'en speed \citep{cassak07b} $v_{Aep}\sim$ 0.5 $v_{te}$, $U_\psi \sim$ 2--4 $v_{Aep}$ is on the order of the electron Alfv\'en speed. $\nabla\cdot\mathbf{U}_\psi$ is $\sim$2 times the electron gyro-frequency $\Omega_{ce}$. These are consistent with the range of $\mathbf{U}_\psi$ from ion to electron Alfv\'en speeds and $\nabla\cdot \mathbf{U}_\psi$ of order 0.1 $\Omega_{ce}$ or higher reported in 2D simulations \citep{Li:2021} and MMS observations \citep{Qi:2022}. In (i), $\delta \mathbf{u}_{e\perp}$ shows an upward electron outflow jet (red arrows) and downward outflows from Xa. In (j), $\delta \mathbf{u}_{i\perp}$ similarly reveals bi-directional ion outflows from the X-line, consistent with reconnection.

How does reconnection spatially distribute in kinetic-scale turbulence? We apply the MFT method in the 3D domain to address this fundamental question.


\begin{figure}
\resizebox{3.85in}{!}{\includegraphics[scale=1.,trim=2.5cm 1.0cm 1.5cm 1.cm, clip=true]{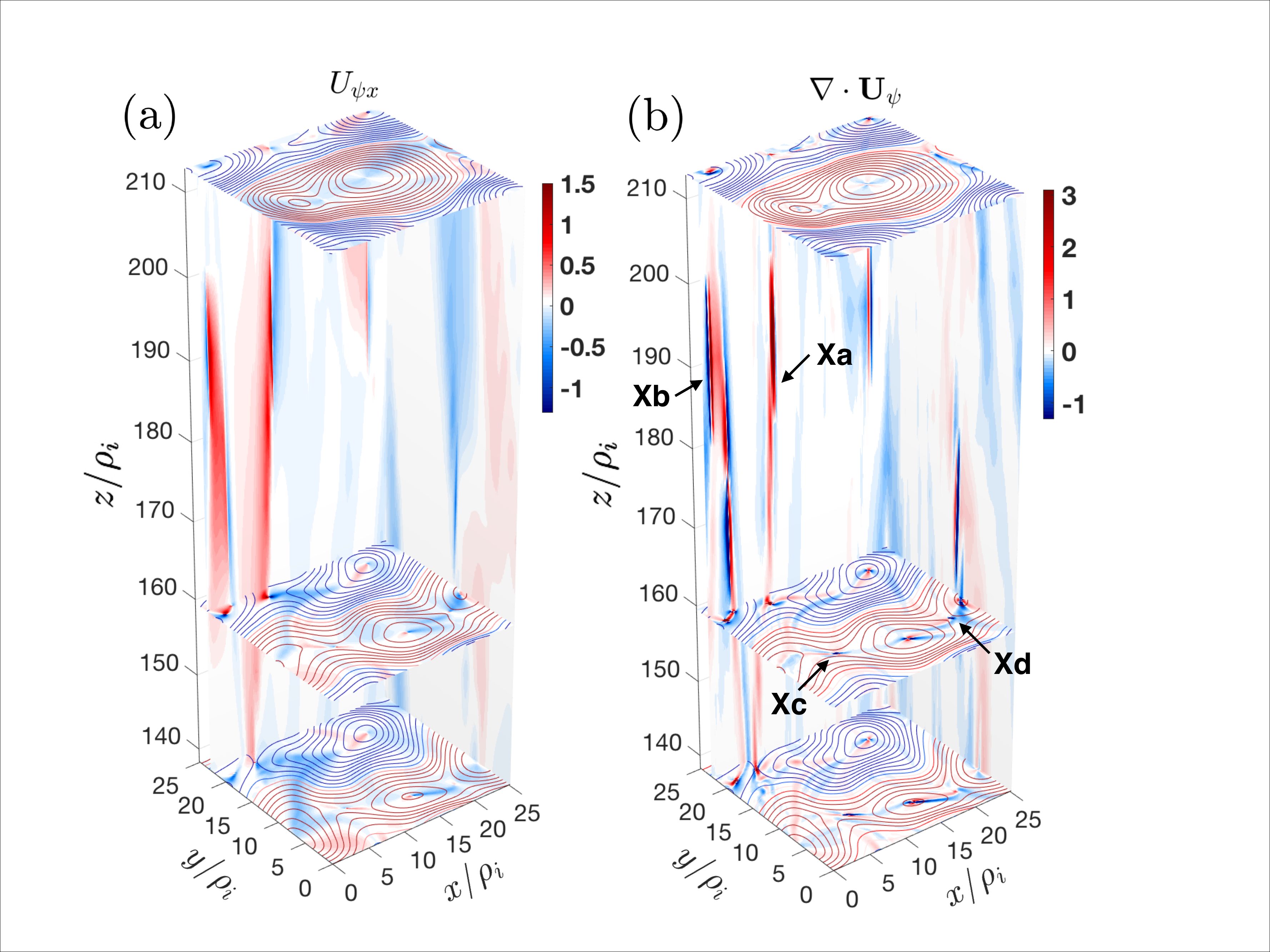}}
\caption{ \label{fig:f3}{\sffamily Extended reconnection: (a) $U_{\psi x}$ and (b) $\nabla\cdot \mathbf{U}_\psi$ reveals extended reconnection X-lines (labeled) in a central region of $z/\rho_i$=140--210. A $xz$ cut through the two strongest X-lines at $y/\rho_i$=23, a $yz$ cut at $x/\rho_i$=25, and $z$-planes at $z/\rho_i$=140, 160 and 210 overlaid with $A_\parallel$ contours, are shown. }}
\end{figure}


{\it Extended reconnection at kinetic scales.} 
Application of MFT to the 3D domain reveals extended reconnection X-lines in kinetic turbulence. We show in \figref{fig:f3} (a) $U_{\psi x}$ and (b) $\nabla\cdot \mathbf{U}_\psi$ for the central region of $z/\rho_i$=140--210. An $xz$ cut at $y/\rho_i$=23 passing through the two strongest X-lines, Xa and Xb, and $z$-planes at $z/\rho_i$=140, 160 and 210, are shown. On the $z/\rho_i$=160 plane, similar to \figref{fig:f1}(d), $U_{\psi x}$ shows converging MFT inflows at Xa, and diverging outflows at Xb. In this 3D region, $U_{\psi x}$ reveals extended inflows at Xa, extending through the entire region from $z/\rho_i$=140 to 210. There is also signature in $\nabla\cdot \mathbf{U}_\psi$ as strong localized positive and negative peaks at Xa, in the form of a quadrupolar structure on the planes of $z/\rho_i$=140 and 160, which extends to $z/\rho_i$=210. Reconnection is highly extended. Here both MFT signatures in $\mathbf{U}_\psi$ and $\nabla\cdot \mathbf{U}_\psi$ are present along the extent of Xa. The same procedure of identification reveals more extended reconnection X-lines in this region, including Xb, Xc, and Xd, as labeled. Supplementary Table (A1) gives the magnitudes of $\mathbf{U}_\psi$ and $\nabla\cdot \mathbf{U}_\psi$ for the X-lines.

We estimate the X-line extents along $z$ from their lower to upper $z$-ends based on MFT signatures, listed in \tabref{tab:extents}. Both reconnection signatures, (i) inflows and outflows in $\mathbf{U}_\psi$ and (ii) strong positive and negative peaks in $\nabla\cdot \mathbf{U}_\psi$, are present along the extent of each reconnection X-line. The X-line extents are of order 100$\rho_i$, which is on the same order of magnitude as the system size $L_z$ = 330$\rho_i$.

\begin{table}[b!]
\centering
\caption{Reconnection X-line extents} \label{tab:extents}
\begin{tabular}{cccc}
\hline
\hline
 X-line & $z_{lower}$ & $z_{upper}$   & Extent ($\rho_i$)  \\
\hline
   Xa &  130   & 210   &  80  \\ 
   Xb &  140   &  200  & 60  \\
   Xc &  110 & 170 & 60 \\ 
   Xd &   130 & 220 & 90 \\ 
\hline
\end{tabular}
\end{table}


How do extended reconnection X-lines develop in kinetic-scale turbulence? We investigate the time evolution of the developing X-lines to address this important question.



\begin{figure}
\resizebox{4.in}{!}{\includegraphics[scale=1.,trim=3.2cm 4.8cm 1.7cm 2.cm, clip=true]{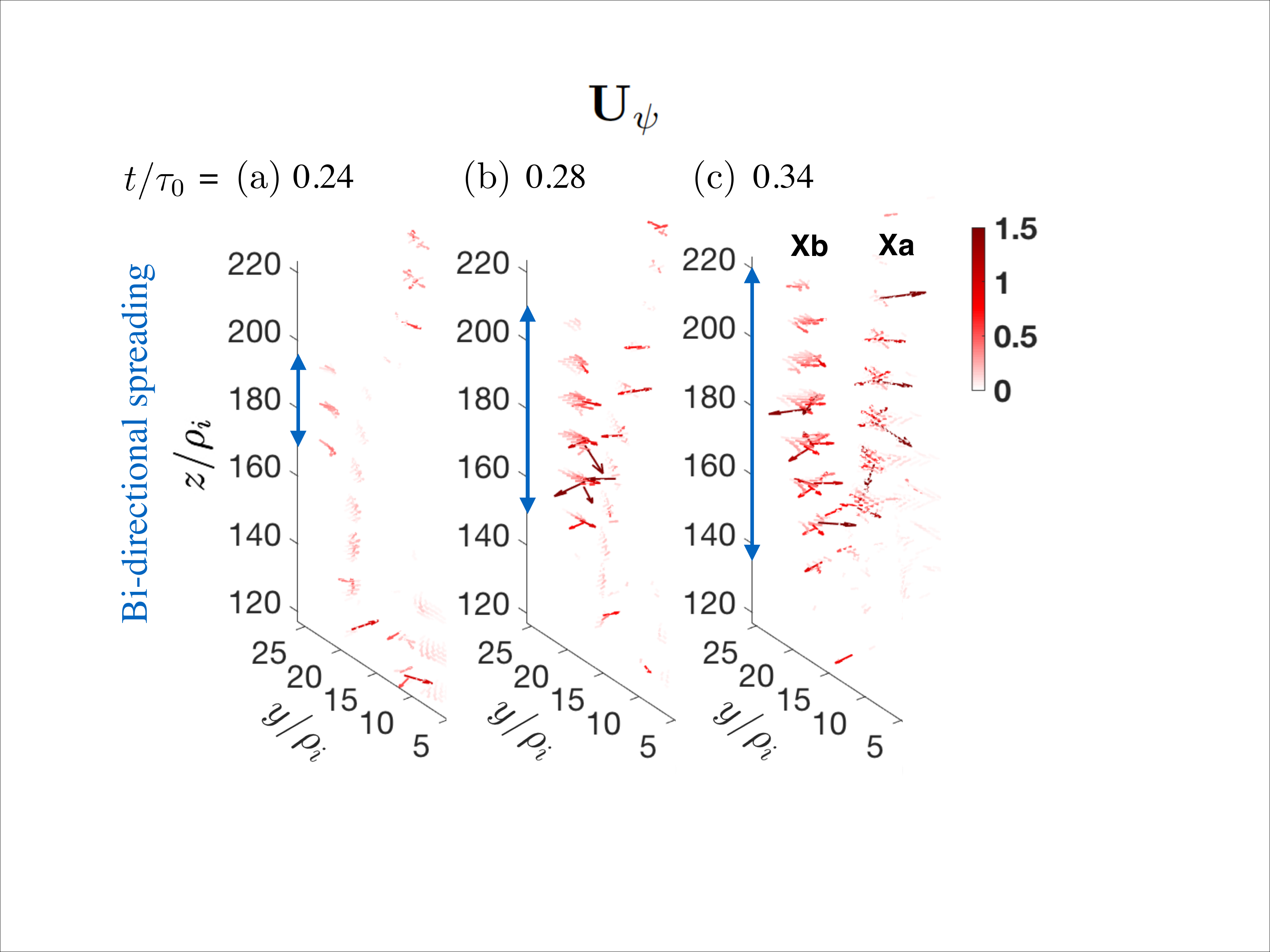}}
\caption{ \label{fig:f5}{\sffamily Time evolution of Xa and Xb showing the development of the X-lines via 3D bi-directional reconnection spreading. Vectors of $\mathbf{U}_\psi$ (similar to \figref{fig:f1}(g)) at three subsequent times, $t/\tau_0$ = (a) 0.24, (b) 0.28 and (c) 0.34, are shown. The $z$ axis is scaled down three times. }}
\end{figure}

{\it Bi-directional reconnection spreading.} 
We show in \figref{fig:f5} the evolution of $\mathbf{U}_\psi$ for a region around the two strongest X-lines at three subsequent times, $t/\tau_0$ = (a) 0.24, (b) 0.28 and (c) 0.34. (a) At $t/\tau_0$ = 0.24, reconnection at Xb arises, weak and localized. (b) At the next time, Xb has extended in the $\pm z$ directions, and also strengthened. Reconnection at Xa has started, with the X-line forming. (c) By $t/\tau_0$ = 0.34, Xb has further extended bi-directionally, and strengthened further; similarly for Xa. Similar evolution is observed for Xc and Xd. The extended X-lines develop via bi-directional reconnection spreading. 

The spreading in the $\pm z$ directions is largely symmetric. The speed of spreading of Xb is estimated to be $\sim v_A$ (see supplementary material, including Refs. \citep{Li:2020a,Shepherd:12}), which is much higher than the electron current speed of $\sim$ 0.5 $v_{te}$ = 0.25 $v_A$ or ion current speed of $< v_{ti}$ = 0.1 $v_A$ (not shown).

This result shows that reconnection that arises in a localized region will develop into a highly extended X-line along the X-line direction through bi-directional spreading. Patchy reconnection with short extents along the X-line direction in laminar sub-ion scale systems \citep{Pyakurel:2021} may be a result of the absence of turbulence driving and/or insufficient time to develop into extended X-lines.


{\it Balance of parallel and perpendicular scales.} We now shed light on what governs the extents of the reconnection X-lines by comparing the parallel and perpendicular time scales of the X-lines.
Recent magnetohydrodynamic (MHD) simulation of merging (reconnecting) flux tubes shows agreement with critical balance \citep{Zhou:2020}, a balance between parallel and perpendicular time scales of fluctuations in anisotropic turbulence \citep{Goldreich:1995}. For extended reconnection X-lines, the parallel time scale is the X-line spreading time that is directly related to its (parallel) extent, which is $\sim$ 0.1 $\tau_0$ for the X-lines. The perpendicular time scale can be based on the inflow speed of reconnection or perpendicular Alfv\'en speed. Considering the strongest X-line, Xa, the perpendicular time scale based on reconnection inflow is $\tau_{R\perp}\sim R/U_{\psi,in}$, where $R\sim$ 5$\rho_i$ is the scale of the reconnecting flux ropes, and $U_{\psi,in}\sim$ 0.2--0.4 $v_{te}$ (Fig. 1(g)) is the upstream MFT inflow speed (which is consistent with the upstream ion inflow speed $\delta u_{i,in}\sim v_{ti}$ = 0.2 $v_{te}$, Fig. 1(j)), giving $\tau_{R\perp}\sim$ 2.5--5 $\Omega_{ci}^{-1}$= 0.1--0.2 $\tau_0$. Alternatively, the time scale based on the perpendicular (upstream) Alfv\'en speed $v_{Ap}$ \citep{cassak07b} is $\tau_{A\perp}\sim R/v_{Ap}$, where $v_{Ap}/v_{ti} = v_{Aep}/v_{te} \sim$ 0.5 (Table (A1)), yielding $\tau_{A\perp}\sim$  0.4 $\tau_0$. The shorter of the two time scales, $\tau_{R\perp}$, is taken as the more dominant perpendicular time scale. The parallel time scale for reconnection, given by the X-line spreading time, $\tau_{R\parallel} \sim$ 0.1 $\tau_0$, approximately balances $\tau_{R\perp}$. Critical balance is satisfied; similarly for Xb and Xd. Reconnection X-lines in kinetic turbulence satisfy critical balance.


{\it Discussion and outlook.} The results presented in this Letter constitute first evidence for extended magnetic reconnection X-lines in kinetic plasma turbulence, and extended X-lines developing through bi-directional reconnection spreading, reaching extents on the same order of magnitude as the system size. This presents a picture that reconnection fundamentally operates in extended regions in kinetic-scale turbulence.


In anisotropic plasma turbulence, the parallel and perpendicular time scales of the fluctuations are balanced by the critical balance relation \citep{Goldreich:1995}. This relation produces anisotropy in both large-scale MHD and small-scale kinetic turbulence, which is observed in numerical simulations at MHD \citep{Shebalin:1982,Cho:2000,Maron:2001,Eyink:2013} and kinetic scales \citep{Cho:2004,Howes:2008a,Cho:2009,TenBarge:2012a,TenBarge:2013b}, including kinetic Alfv\'en and whistler turbulence. Not only the turbulent fluctuations, but reconnection in turbulence also satisfies critical balance, evident in MHD simulations \citep{Zhou:2020} and our gyrokinetic simulation, which produces extended reconnection X-lines.
This implies that reconnection X-lines are coherent structures, with their parallel and perpendicular scales related to each other. This relation provides a way to predict the extent of reconnection X-lines at a given perpendicular scale, confirming that reconnection X-lines will be highly extended at kinetic scales (where $\delta B_\perp \ll B_0$). For reconnection X-lines observed at large scales with an extent over 10$^4\rho_i$, assuming order one fluctuations ($\delta B_\perp \sim B_0$), the perpendicular scales of the associated ICMEs are predicted to be similarly over 10$^4\rho_i$, which is consistent with statistical observations near Earth \citep{Cane:2003}.

The tearing instability is one of the instabilities known to be important for driving reconnection in plasmas. Reconnection in our kinetic simulation does not appear to be driven by the tearing instability, which is consistent with the lack of tearing-driven reconnection in simulations of turbulent reconnection at MHD scales \citep{Oishi:2015,Beresnyak:2017,Kowal:2020}. This supports the similarity of reconnection in turbulence across scales.

At $k_\perp\rho_i>$1, the gyrokinetic model used here describes kinetic Alfv\'en wave turbulence that satisfies $k_\parallel\ll k_\perp$ and critical balance; although in the low-frequency limit (below the ion gyro-frequency), it is consistent with 3D fully kinetic simulations that retain high-frequency fluctuations \citep{Grovselj:2018} and solar wind observations \citep{Howes:2008a}. The results presented here are expected to hold more generally in fully kinetic plasmas.

Numerous studies have examined the general problem of reconnection in magnetized turbulence in the MHD limit \citep{Lazarian:2020}. For instance, the level of MHD turbulence is found to be important for determining the reconnection rate in 3D \citep{Lazarian:1999,Kowal:2009}. Reconnection in MHD turbulence may share similarities with that in kinetic turbulence studied here. A detailed analysis of the reconnection rate and comparison with previous work, while beyond the scope of the current work, promises future work.

With applicability to both simulations and observations \citep{Qi:2022,Qi:2023,Wang:2023}, the MFT method opens opportunities for studying reconnection in turbulence. Although here we have identified extended reconnection in kinetic turbulence, future work could explore how extended X-lines contribute to plasma heating at kinetic scales, how reconnection spatially distributes in electron-scale turbulence, and how properties of reconnection change with turbulent conditions in space, astrophysical, and laboratory plasmas.



\begin{acknowledgments}
The authors thank N. Loureiro and Y. D. Yoon for helpful discussions. This work is supported by NSF award AGS-2000222 and NASA MMS mission NNG04EB99C. It used the Extreme Science and Engineering Discovery Environment (XSEDE), which was supported by NSF award ACI-1053575.
\end{acknowledgments}



\providecommand{\noopsort}[1]{}\providecommand{\singleletter}[1]{#1}%
%

\section{Supplementary Material}\label{sec:supp}
This section includes a comparison of $\mathbf{U}_\psi$ and the E$\times$B velocity, information on the simulation code and setup, a table on the magnitude of the MFT quantities and an estimate of reconnection spreading speed.

{\it Comparison of $\mathbf{U}_\psi$ and E$\times$B velocity.} The fluid velocity in ideal MHD (i.e., $\mathbf{E} + \mathbf{v}\times \mathbf{B}$=0) can be considered as the magnetic field line velocity because this velocity is flux preserving \citep{Vasyliunas:1972}. This field line velocity equals the E$\times$B velocity when $\mathbf{E}\cdot\mathbf{B}$ = 0, but this equivalence fails if $\mathbf{E}\cdot\mathbf{B}\ne$ 0 \citep{Pontin:2022}. Derived from the advection equation of magnetic flux, $\mathbf{U}_\psi$ does not have these constraints, and is applicable to non-ideal and ideal plasmas alike. In the ideal limit of $\mathbf{E}\cdot\mathbf{B}$ = 0 ($\mathbf{B}_p, E_z \ne$ 0 and $\mathbf{E}_p, B_z$= 0), $\mathbf{U}_\psi$ equals the E$\times$B velocity.


{\it Simulation code.} The simulation was performed \citep{Li:2016} using \T{AstroGK} \citep{Numata:2010}. \T{AstroGK} has been extensively used to investigate turbulence in weakly collisional plasmas \citep{Howes:2008a,Howes:2011a,TenBarge:2012a,TenBarge:2013a,Li:2016,Li:2019}.
\T{AstroGK} is an Eulerian continuum code with triply periodic
boundary conditions. It has a slab geometry elongated along the
straight, uniform background magnetic field, $\V{B}_0=B_0 \zhat$. The
code evolves the perturbed gyroaveraged Vlasov-Maxwell equations in
five-dimensional phase space (three-dimensional-two-velocity)
\citep{Frieman:1982,Howes:2006}. The evolved quantities are the
electromagnetic gyroaveraged complementary distribution function
for each species $s$, the scalar
potential $\varphi$, parallel vector potential $A_\parallel$ and
parallel magnetic field perturbation $\delta B_\parallel$, where
$\parallel$ is along the total local magnetic field $\V{B}=B_0\zhat+
\delta \V{B}$. The total and background magnetic fields are the same, 
to first-order accuracy, which is retained for perturbed fields in gyrokinetics.
The velocity grid is specified by pitch angle $\lambda=v_\perp^2/v^2$ and 
energy $\varepsilon=v^2/2$. 
The background distribution functions for both species are stationary
uniform Maxwellians. Collisions are incorporated using a fully
conservative, linearized gyro-averaged Landau collision operator \citep{Abel:2008,Barnes:2009}.

{\it Setup.} We specifiy here a 3D generalization \citep{Li:2016,Li:2019} of the classic 2D Orszag-Tang Vortex (OTV) problem \citep{Orszag:1979}. The 2D problem was widely used in fluid and magnetohydrodynamic turbulence simulations; various 3D generalizations 
have also been used for studying turbulence \citep{Politano:1989,Mininni:2006,Parashar:2014b}
This 3D OTV setup consists of counterpropagating \Alfven waves along
$\V{B}=B_0\zhat$ such that on the $z=0$ plane, its initial condition
reduces to that of the 2D OTV problem. An initial amplitude
$\mbox{z}_0$ of Els\"asser variables in the OTV setup \citep{Li:2016}
is chosen to yield a nonlinearity parameter $\chi = k_\perp\mbox{z}_0/(k_\parallel v_A) = 1$ (where $v_A=B_0/\sqrt{4\pi m_in_0}$ is a characteristic \Alfven speed), corresponding to a state of strong turbulence, satisfying \emph{critical balance} \citep{Goldreich:1995}. Note that previous studies using \T{AstroGK}
have shown consistency with the prediction of a critically balanced cascade 
in the dissipation range \citep{TenBarge:2012a,TenBarge:2013b}.

To follow the turbulent cascade from the inertial range ($k_\perp\rho_i \ll 1$) to below electron scales ($k_\perp\rho_e >1$) \citep{TenBarge:2013a,TenBarge:2013b}, we specify a reduced mass ratio, $m_i/m_e=25$, which, in a simulation domain of $L_{\perp}=8\pi \rho_i$ and dimensions $(n_x,n_y,n_z,n_\lambda,n_\varepsilon)=(128,128,32,64,32)$, enables us to resolve a dynamic range of $0.25 \le k_\perp\rho_i \le 10.5$, or $0.05 \le k_\perp\rho_e \le 2.1$. 
Plasma parameters are ion plasma $\beta_i = 8 \pi n_i T_{0i}/B_0^2=0.01$ and $T_{0i}/T_{0e}=1$. Collision frequencies of $\nu_i$ =
10$^{-5} \omega_{A0}$ and $\nu_e$ = 0.05 $\omega_{A0}$ (where
$\omega_{A0} \equiv k_{\parallel}v_A$ is a characteristic Alfv\'en wave
frequency in 3D) are sufficient to keep velocity space well resolved
\citep{Howes:2008a,Howes:2011a}.

\begin{table}[tbh]
\renewcommand{\thetable}{A1}
\centering
\caption{Magnitude of $U_{\psi}$ and $\nabla\cdot \mathbf{U}_\psi$} \label{tab:caep}
\begin{tabular}{cccc}
\hline
\hline
 X-line  &$v_{Aep}$  &  $U_{\psi ,max}$   & $|\nabla\cdot \mathbf{U}_\psi|_{max}$ \\
\hline
   Xa &  0.5  & 2   &  2  \\ 
   Xb &  0.7  & 2   &  2.5  \\
   Xc &  0.7  & 0.3 &  0.4 \\ 
   Xd &  0.4  & 0.5 &  0.6 \\ 
\hline
\end{tabular}
\end{table}

{\it Magnitude of MFT velocity and divergence.} \tabref{tab:caep} lists the upstream electron Alfv\'en speed $v_{Aep}$ \citep{cassak07b}, maximum $U_\psi$ and $|\nabla\cdot \mathbf{U}_\psi|$ of the reconnection X-lines at $z/\rho_i$=160. 
Here the range of $U_\psi$ is on the order of the ion to electron Alfv\'en speeds, and the divergence is of order 0.1 $\Omega_{ce}$ or higher, consistent with 2D simulation \citep{Li:2021} and MMS observations \citep{Qi:2022}.


{\it Speed of reconnection spreading.} Here we estimate the speed of reconnection spreading along $z$. Considering one of the strongest X-lines, Xb, the X-line spreads half of its extent, 30$\rho_i$ (Table (I)), in a time scale of $\simeq$ 0.1$\tau_0$ = 2.5$\Omega_{ci}^{-1}$. This yields a spreading speed of $v_{sp}\simeq 30\rho_i/2.5 \Omega_{ci}^{-1}$ = 1.2 $v_A\simeq v_A$. The spreading speed of the X-line agrees with the phase speed of kinetic Alfv\'en waves constituting the turbulence in the system at kinetic scales, consistent with the waves bi-directionally spreading the X-line. This is consistent with spreading at $v_A$ in laminar reconnection simulations without turbulence \citep{Li:2020a,Shepherd:12}.


\end{document}